\documentclass[preprint,showpacs,preprintnumbers,amsmath,amssymb]{revtex4}

\usepackage{graphicx}
\usepackage{dcolumn}
\usepackage{bm}
\usepackage{amsmath}
\graphicspath{{Figure/}}
\newcommand{\DT}{\Delta \theta}

\begin{document}

\title{Minimal Model for Synchronization Induced by Hydrodynamic Interactions}
\author{Bian Qian}
\author{Hongyuan Jiang}
\author{David A. Gagnon}
\author{Kenneth S. Breuer}
\email{kbreuer@brown.edu}
\author{Thomas R. Powers}
\email{thomas_powers@brown.edu}
\affiliation{Division of Engineering, Brown University}
\date{April 15, 2009; revised December 4, 2009}

\begin{abstract}
Motivated by the observed coordination of nearby beating cilia, we use a scale model experiment to show that hydrodynamic interactions can cause synchronization between rotating paddles driven at constant torque in a very viscous fluid. Synchronization is only observed when the shafts supporting the paddles have some flexibility. The phase difference in the synchronized state depends on the symmetry of the paddles. We use the method of regularized stokeslets to model the paddles and find excellent agreement with the experimental observations. We also use a simple analytic theory based on far-field approximations to derive scaling laws for the synchronization time as a function of paddle separation.

\end{abstract}


\maketitle

\section{Introduction}

One of the central aims in the field of cell motility is to understand how a collection of beating cilia coordinates, or, on a larger scale, how a collection of swimming organisms form coherent patterns. For example,  \textit{Paramecium} swims by propagating waves of ciliary beating along its surface~\cite{machemer1972}. The alga \textit{Chlamydomonas} beats its two flagella in synchrony to swim straight and asynchronously to change its orientation~\cite{PolinTuvalDrescherGollubGoldstein2009,GoldsteinPolinTuval2009}.
At the level of a population of cells, sea urchin spermatozoa spontaneously form vortex patterns in the absence of cell signaling~\cite{reidel}.
Coordination of cilia is also important in the transport of fluid. The coordination of nodal cilia in developing vertebrate embryos has been implicated in the determination of left-right asymmetry of the organism~\cite{nonaka_etal1998}. The cilia lining the human airway must beat in a coordinated manner to sweep foreign particles up the airway. Beating cilia may also play a role in the transport of sperm and egg during fertilization in mammals~\cite{suarez06}.

These examples are instances of the general tendency for the emergence of synchronization in a broad array of physical and biological systems~\cite{PikovskyRosenbluKurths2001}. In this article we investigate the long-standing hypothesis that the coordination observed in nearby beating cilia or swimmers is due to hydrodynamic interactions between these objects~\cite{gray1928,Sleigh1974}. In recent years there have been many computational and theoretical studies to support this hypothesis~\cite{fauci90, gueron97, lagomarsino03, lagomarsino02,reichert, lenz06,vilfan, guirao07,niedermayer08,ElfringLauga2009}. The key physical fact underlying all of these studies is that at the small scale of the cell, where the Reynolds number $\mathrm{Re}\ll1$,  the velocity field arising from a deforming body falls off slowly with distance, leading to significant hydrodynamic forces between nearby bodies.  Furthermore, the development of a fixed phase difference between two bodies---phase-locking---requires some kind of compliance in which the deforming body can adjust its beat pattern in response to hydrodynamic forces from other nearby bodies.

The nature of this compliance is subtle. In the case of two rotating rigid helices driven with fixed torques (the ``deformation" here is rotation), the freedom of the phase of each helix to speed up or slow down to maintain the fixed torque for all phase differences does \textit{not} lead to phase-locking~\cite{kim}. Theoretical calculations suggest that additional degrees of freedom are required for phase-locking, or synchronization. For example, synchronization develops if the shafts of the rotating helices are connected to fixed points by stiff springs, allowing the axes of the helices to translate or tilt~\cite{reichert}. The directions of these small motions depend on whether the hydrodynamic forces are attractive or repulsive, which in turn depends in detail on the phase difference (cf. the case of nearby swimmers~\cite{alexander08,laugabartolo08}).

\begin{figure}
\includegraphics[width=3.1in]{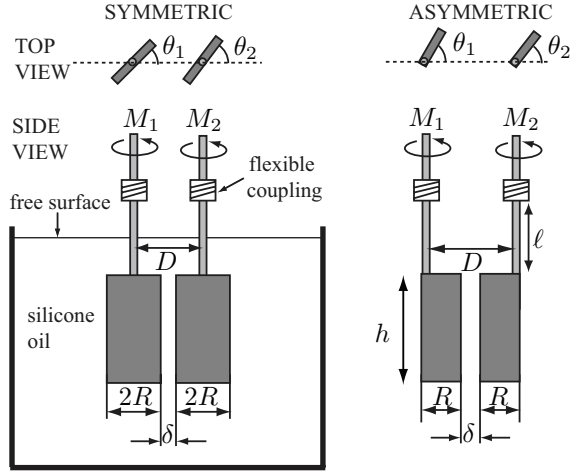}
\caption{Schematic of the model system for hydrodynamic synchronization. Left: A pair of symmetric paddles 
in a fluid with viscosity $\eta$ are rotated with constant torques $M_1$ and $M_2$.  
The  shafts are rigid but have flexible couplings that allow the paddles to tilt. Right: asymmetric paddles.}
\label{fig:setup}
\end{figure}

The complexities of designing experiments that include both hydrodynamic interactions and controlled elastic deformation at very low Reynolds numbers have hindered experimental studies of hydrodynamic synchronization; therefore, we built a scale model system that captures the essential physics, allows for detailed measurements, and is amenable to modeling.  This article presents results from experiments (\S\ref{expsec}), numerical simulations (\S\ref{numersec}), and a theoretical model (\S\ref{asymsec} and \S\ref{symsec}) that together outline a coherent framework for describing hydrodynamic synchronization.

\section{Experiment}
\label{expsec}

Figure~\ref{fig:setup} illustrates the experimental configuration.  Two thin paddles are immersed in a large tank ($60 \times 60 \times 60$\,cm) filled with a viscous fluid ($\eta = 110$\,Ns/m$^2$), separated at their closest approach by a small gap, $\delta=3.6$\,mm.  We study two different paddle configurations: symmetric and asymmetric. The symmetric paddle has the axis of rotation through the paddle center and dimensions $h=60$\,mm, $w=2R=30$\,mm, and thickness $t=6$\,mm. The asymmetric paddle has the axis of rotation through one edge and dimensions of $60 \times 20 \times  6$\,mm.  
The paddles are small compared to the size of the tank. By repeating some of the experiments with the paddles at different positions within the tank, we confirmed that the side walls did not affect the results in any appreciable manner.

The paddles are supported by  shafts  that are  hardened steel, of diameter 6.35\,mm and length $\ell=120$\,mm, connected to the motors via flexible couplings that allow the paddles to tilt. The shafts are so rigid that bending due to hydrodynamic forces is negligible, but the couplings act as torsional springs with spring constant  $k_\mathrm{T}=8000$\,mN-m/rad, leading to an equivalent spring constant for lateral shifts of the paddles of $k=k_\mathrm{T}/\ell^2$. This flexibility allows the paddles to tilt slightly in response to hydrodynamic forces. We also tested shafts without an intermediate coupling, in which the ability of the paddles to tilt effectively vanished.
The bearing assemblies are supported on separate stages to minimize any mechanical communication beyond hydrodynamic interactions~\cite{Huygenss-Clocks}, and to allow for precise control of the distance of closest approach, $\delta$. Since $\delta/h\ll1$, the resultant flow is mostly two-dimensional, in the plane perpendicular to the axes of rotation.

The two paddles are driven at constant torque using a DC servo motor, digital encoder, load cell, and feedback controller. Each paddle is driven by a servo motor which is encased in a housing. To measure the torque delivered by the motor, the housing is supported by bearings and prevented from rotating by a rigid, $\approx10$\,cm-long torque arm.  Due to the bearings, the entire reaction torque on the housing is transmitted by the torque arm to a precision load cell.
The load cell output signal is used as a feedback to a PID controller that adjusts the voltage driving the servo motor, thus maintaining a defined torque.  The PID controller updates at approximately 100\,Hz---500 times faster than the typical rotational frequency of the paddles in the experiment (0.2\,Hz). The position of the paddle is recorded from the output of the digital encoder at each update of the PID controller.  Velocity is calculated from the position using high-order finite differences. The system was calibrated by measuring the rotational speed vs. voltage for an isolated paddle over a range of torques and using the theoretically-known torque-speed relationship to associate the measured load cell voltage with a specific torque. The accuracy and stability of the system was verified by measuring (i) the torque fluctuation for an isolated paddle rotating at constant speed, and (ii) the velocity fluctuation of an isolated paddle rotating at constant torque.  In both configurations, we confirmed that the system was stable to better than 1.5\% of the set point.  Typical driving torques range from 4\,mN-m to 25\,mN-m, corresponding to rotation frequencies no more than 0.2\,Hz.  
\begin{figure}
\includegraphics[width=3.1in]{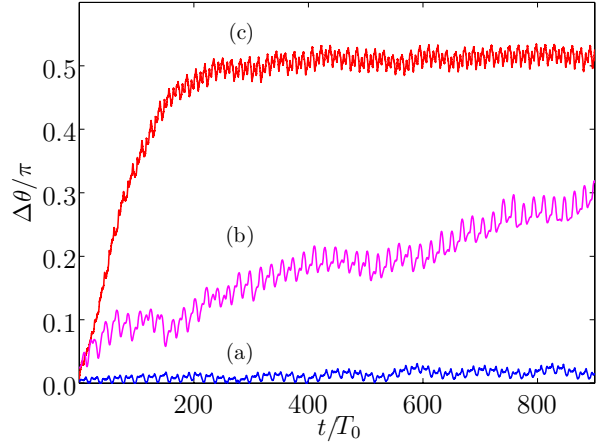}
\caption{(Color online.) Phase difference $\Delta\theta = \theta_2 - \theta_1$ vs. dimensionless time $t/T_0$ ($T_0=6\pi\eta R^3/M$) for symmetric paddles with (a) $M_1=M_2$ and stiff shafts, (b) $(M_2-M_1)/M_1\approx0.003$ and stiff shafts, and (c) $M_1=M_2$ and shafts with flexible couplers. }
\label{fig:stiff_flexible}
\end{figure}

\begin{figure}
\includegraphics[width=3.2in]{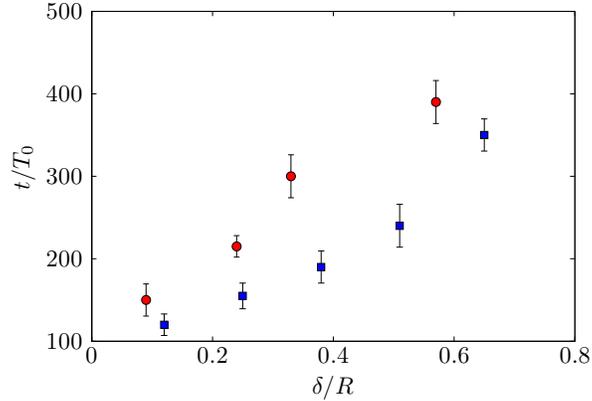}
\caption{(Color online.) Dimensionless synchronization time vs. dimensionless gap size $\delta/R$ for symmetric paddles (circles) and asymmetric paddles (squares). For symmetric paddles, the dimensionless synchronization time is measured from the moment of phase difference $\Delta \theta = 0.1$ to the time of the first stable state $\Delta \theta = \pi/2$. For asymmetric paddles, the dimensionless synchronization time is defined as the time from $\Delta \theta = 0.8$ to $\Delta \theta = 0$. The uncertainty comes from the low frequency fluctuation in $\Delta \theta$ due to the system noise. Time is normalized by $T_0 = 6 \pi \eta R^3/M$.}
\label{fig:gap_syn}
\end{figure}

\begin{figure}
\includegraphics[width=3.2in]{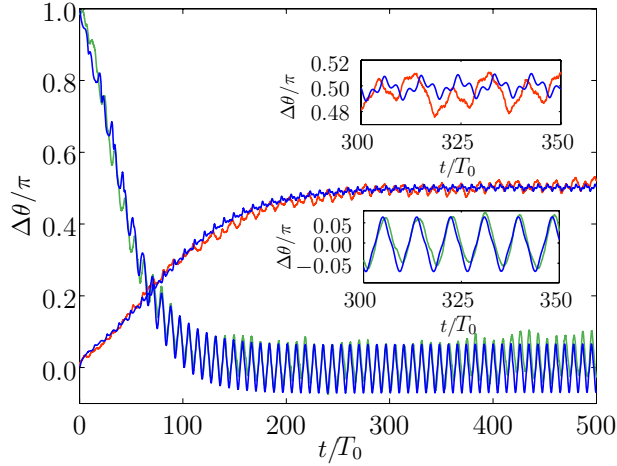}
\caption{(Color online.) Phase difference vs. dimensionless time for symmetric (red line) and asymmetric paddles (green line), compared to simulation (blue lines). Time is measured in units of $T_0=6\pi\eta R^3/M$.
The insets show the phase difference once phase-locking is achieved.  In both cases, the normalized gap 
between the paddles is $\delta /R=0.24$. }
\label{fig:syn}
\end{figure}

\begin{figure}
\includegraphics[width=3.1in]{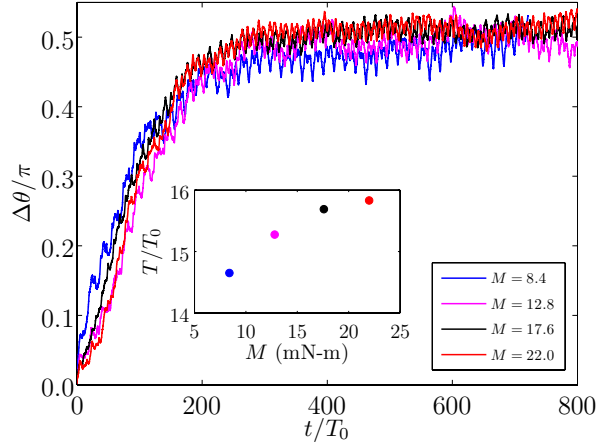}
\caption{(Color online.) Phase difference vs. dimensionless time $t/T_0$ for symmetric paddles driven at  torques $M_1=M_2=8.4, 12.8, 17.5,$ and $22$\,mN-m, with $\delta /R=0.24$. The collapse of the data shows that the time to synchronize scales with $T_0=6\pi\eta R^3/M$. The inset shows the dimensionless period $T/T_0$ in the synchronized state.}
\label{fig:scaletor}
\end{figure}

At these conditions, the Reynolds number,  $\mathrm{Re}= \rho \omega R^2/\eta\approx10^{-3}$, is small enough to justify the neglect of inertial forces.  This was confirmed experimentally by noting that if the paddle rotation was initiated with a constant velocity, the time taken to reach constant torque was less than 250\,ms. For these Stokes flows, the characteristic velocities scale linearly with the motor torques $(M_1, M_2)$, and the state of the system is determined by  the angles of the two paddles $(\theta_1,\theta_2)$ (Fig.~\ref{fig:setup}) and the small shifts of the paddles due to the flexible couplers. In the high-stiffness case, the paddles did not synchronize in any measurable time; instead, the phase of each paddle increased roughly linearly with driving torque, $(\theta_1,\theta_2)\propto (t M_1,t M_2)$, independent of the initial phase difference (Fig.~\ref{fig:stiff_flexible}a, b). However, paddles with flexible couplers and small distance of closest approach locked phases in $10$--$20$ revolutions (Fig.~\ref{fig:stiff_flexible}c).  Note that we measure time in units of $T_0 = 6\pi\eta R^3/M$, ($M$ is the mean torque) which is roughly one tenth of a rotation period.  The data we display in this article is for a dimensionless gap size $\delta/R=0.24$. We also varied $\delta/R$ for both kinds of paddles from $\approx0.1$ to $\approx0.6$, and found that the time to synchronize increased with spacing, with longer times and a faster increase for the symmetric paddles (Fig.~\ref{fig:gap_syn}).

For $M_1 = M_2$, the symmetric paddles  locked phases at $\DT\equiv\theta_2-\theta_1= \pi/2$, and the asymmetric paddles settled at $\DT = 0$  (Fig.~\ref{fig:syn}).  These two states represent the conditions that roughly maximize the distance of closest approach of the two paddles.  Since the paddles would minimize their distance of closest approach  if they maintained their typical initial phase differences ($\DT = 0$ for the symmetric paddles, $\DT = \pi$ for the asymmetric paddles), the rotation speed of each paddle rises as the paddles synchronize. Denoting the rotation speed of an isolated paddle by $\omega_0$, we found that the speed of both symmetric paddles rises from $0.72\omega_0$ to $0.85\omega_0$ as synchronization develops, whereas the speed of both asymmetric paddles  rises from $0.75\omega_0$ to $0.93\omega_0$. While these synchronized states are stable, there is a consistent and repeatable phase fluctuation (Fig.~\ref{fig:syn}-inset) corresponding to the variation in rotational speeds as the hydrodynamic interactions between the paddles wax and wane during a cycle.  The fluctuation amplitude in the asymmetric case is larger than in the symmetric case because there is a larger variation in the distance between the asymmetric paddles during a period. These observations qualitatively agree with the results of numerical calculations on rotating rigid helices with flexible couplers~\cite{reichert}. In our experiments the phase fluctuations and rise in velocity as synchronization develops are more dramatic since the variation in the hydrodynamic interaction between paddles over a period is greater than in the case of helices.

\begin{figure}
\includegraphics[width=3.2in]{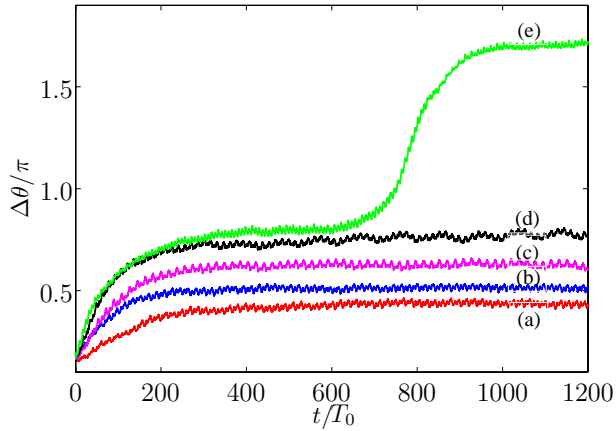}
\caption{(Color online.) Phase difference vs. dimensionless time for symmetric paddles with $\delta /R=0.24$ and $\Delta M/M_1=(M_2-M_1)/M_1=-1\%$ (a), $\Delta M/M_1=0\%$ (b), $\Delta M/M_1=1\%$ (c), and $\Delta M/M_1=2\%$ (d). When $\Delta M/M_1\approx3\%$, the synchronized states become unstable, and the system exhibits transition in which $\Delta\theta$ jumps by $\pi$.
}
\label{fig:tordif}
\end{figure}

The final state of synchronization was found to be independent  of the initial orientation of the paddles. The time to synchronize scales with $T_0$, perhaps with a weak dependence on torque (Fig.~\ref{fig:scaletor}). The number of paddle revolutions needed to synchronize is therefore roughly constant, 15 in the case of symmetric paddles, and $20$ for asymmetric paddles. In the synchronized state, however, the dimensionless rotation period $T/T_0$ increases slightly with torque (Fig.~\ref{fig:scaletor}, inset).  When the symmetric paddles are operated with a torque mismatch between the two motors, the synchronized phase difference increases with $\Delta M$, although for a large mismatch, $\Delta M/M_1=3\%$, the synchronized state is only marginally stable and the phase difference can jump abruptly by $\DT=\pi$ (Fig.~\ref{fig:tordif}).

\section{Numerical Simulations}
\label{numersec}

These experiments give strong evidence that the phase-locking of the paddles is due to hydrodynamic interactions. We tested this hypothesis by using the method of regularized Stokeslets~\cite{Cortez2001} to model the flows induced by the paddles. Each paddle is replaced by a rectangular array of regularized Stokeslets $S^b_{\mu\nu}$ with strength $f_\nu$, where $\mu$ and $\nu$ label the Cartesian coordinates $x$, $y$, and $z$. The flow from the Stokeslet at $\mathbf{x}'$ is given by
\begin{equation}
v_\mu(\mathbf{x})=\sum_{\nu}S^b_{\mu\nu}(\mathbf{x},\mathbf{x}')f_\nu,
\end{equation}
with associated pressure
\begin{equation}
p(\mathbf{x})=\sum_{\nu}\frac{p^b_\nu(\mathbf{x},\mathbf{x}')f_\nu}{8\pi}.
\end{equation}
The Stokeslet $S^b_{\mu\nu}$ satisfies
\begin{eqnarray}
\sum_{\nu}\partial S^b_{\mu\nu}/\partial x_\nu &=& 0  \\
\nabla^2S^b_{\mu\nu}(\mathbf{x},\mathbf{x}')-{\partial p^b_\nu}/{\partial x_\mu} &=& -8\pi\delta_{\mu\nu}\phi_b(\mathbf{x}-\mathbf{x}'),
\end{eqnarray}
where $\phi_b(\mathbf{x}-\mathbf{x}')$ is a smooth approximation to the Dirac delta function with spread $b$,
\begin{equation}
\phi_b(\mathbf{x}-\mathbf{x}')=\frac{15 b^4}{8\pi\left(r^2+\epsilon^2\right)^{7/2}},
\end{equation}
and $r=|\mathbf{x}-\mathbf{x}'|$.
The number of stokeslets and the spread $b$ are chosen to give good agreement between the measured and simulated resistance coefficient for a single rotating paddle at the center of the tank. The spread $b$ is large enough to make the regularized stokeslets overlap, which prevents fluid from leaking through the paddles. We model the flexibility of the couplers with springs of torsional spring constant $k_\mathrm{T}$. For simplicity we suppose that the shafts are always vertical, but can undergo slight shifts in the horizontal plane.  With the assumption that the paddles are rigid, the degrees of freedom are the angles $(\theta_1, \theta_2)$ of the paddles and the positions of the shafts. Balancing forces and torques leads to coupled nonlinear differential equations which we solve numerically. Figure~\ref{fig:syn} shows the excellent agreement between the experiments and the simulations for both the asymmetric and the symmetric paddles. The simulation accurately captures the frequency and amplitude of the oscillations associated with the rotation of the motors, as well as the slower evolution of the phase-locking. When the driving torque is varied over the range used in the experiment, the simulations yield that the dimensionless time to synchronize  $T_\mathrm{s}/T_0$ remains approximately constant, with a weak dependence on torque, in accord with Fig.~\ref{fig:scaletor}.  Simulations with infinite spring constant $k_\mathrm{T}$ show no phase-locking. Since the paddles in the simulation are coupled only through the hydrodynamic interaction, we conclude that the cause of the phase-locking is the hydrodynamic interaction and not any stray mechanical coupling that might be present in the experimental apparatus.

\section{Simple Model for Asymmetric Paddles}
\label{asymsec}

\subsection{Oseen tensor model}
\begin{figure}
\includegraphics[width=3.3in]{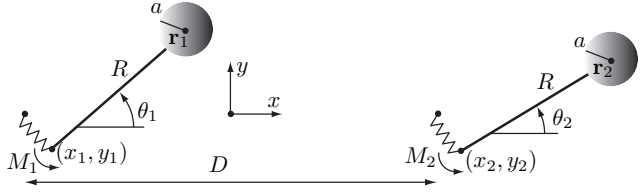}
\caption{Model for asymmetric paddles. The figure is not to scale; note that $a\ll R\ll D$.}
\label{fig:oneball}
\end{figure}

We can gain more insight into the mechanism of phase-locking by developing a simple theory along the lines of reference~\cite{niedermayer08}. A minimal model for the asymmetric paddles is to replace each paddle with a sphere of radius $a$ attached to one end of a  rod of length $R$ (Fig.~\ref{fig:oneball}). The rod is rigid and does not disturb the fluid. The other end of the rod  is attached to a stationary point by a  spring with spring constant $k$. The rods are rotated by moments $M_1$ and $M_2$ which are applied at the ends of the rods attached to the springs, where we can imagine shafts perpendicular to the plane of the page.  The spring is stiff, with $k\gg M_1/R^2$. Denote the positions of the balls by $\mathbf{r}_i=\mp(D/2)\hat{\mathbf{x}}+\mathbf{x}_i+R\hat{\bm{\rho}}_i$, where the minus sign applies for $i=1$, the plus sign applies for $i=2$, and $\hat{\bm{\rho}}_i=(\cos\theta_i,\sin\theta_i)$. Note that $\theta_i$ is defined as the angle the rod makes with the $x$-axis, not the angle $\mathbf{r}_i$ makes with the $x$-axis.
The vectors $\mathbf{x}_1=(x_1,y_1)$ and $\mathbf{x}_2=(x_2,y_2)$ are the displacements of the shafts from the stationary points $(-D/2,0)$ and $(D/2,0)$, respectively.
If we suppose the balls are far apart, with $D\gg a$, then the leading-order interaction  between the two balls is given by the Oseen tensor~\cite{doi_edwards1986}: 
\begin{eqnarray}
\mathbf{v}_1&=&\frac{\mathbf{f}_1}{6\pi\eta a} + \frac{1}{8\pi\eta}\left[\frac{\mathbf{f}_2}{|\mathbf{r}_{12}|}+ \frac{(\mathbf{f}_2\cdot\mathbf{r}_{12})\mathbf{r}_{12}}{|\mathbf{r}_{12}|^3}\right]\\
\mathbf{v}_2&=&\frac{\mathbf{f}_2}{6\pi\eta a} + \frac{1}{8\pi\eta}\left[\frac{\mathbf{f}_1}{|\mathbf{r}_{12}|}+ \frac{(\mathbf{f}_1\cdot\mathbf{r}_{12})\mathbf{r}_{12}}{|\mathbf{r}_{12}|^3}\right],\label{Oseenmodel}
\end{eqnarray}
where $\mathbf{v}_i=\mathrm{d}\mathbf{r}_i/\mathrm{d}t=\dot{\mathbf{r}}_i$ is the velocity of the $i$th ball, $\mathbf{f}_1$ and $\mathbf{f}_2$ are the forces exerted by the balls on the fluid, and $\mathbf{r}_{12}=\mathbf{r}_1-\mathbf{r}_2$.  

Since the spring is assumed linear and the motion of the rod incurs no drag force, the balance of forces on each paddle is $-\mathbf{f}_i-k\mathbf{x}_i=\mathbf{0}$. We must also enforce moment balance. Since inertia is unimportant at $\mathrm{Re}=0$, we may compute moments about the points $\mathbf{x}_i$ for each paddle:
\begin{equation}
M_i+\hat{\mathbf{z}}\cdot
\left(R\hat{\bm{\rho}}_i\right)\times\left(-\mathbf{f}_{i}\right)=0.\label{etorquebal}
\end{equation}

\subsection{Separation of time scales}
The analysis of the equations of motion is simplified by the recognition that our problem has three well-separated time scales: (i) a short time scale $T_k=\eta a/k$ (recall $k=k_\mathrm{T}/\ell^2$) that controls the rate of relaxation of the springs, (ii) an intermediate time scale $T_1=6\pi\eta aR^2/M_1$~\footnote{Note that the period of an isolated paddle of width $R$ (as in Fig.~\ref{fig:setup}) scales as $T_0=6\pi\eta R^3/M_1$, whereas $T_1=6\pi\eta aR^2/M_1$ is a more appropriate scale for the period of a ball of radius $a$ moving on an orbit of approximate radius $R$ and driven by torque $M_1$.} that controls the period of an isolated ball driven by torque $M_1$, and (iii) a long time scale $T_\mathrm{s}$ that characterizes the time for phase-locking to develop.  Since the springs are stiff, $T_k\ll T_1$. Since the interaction between the paddles is weak, $T_1\ll T_\mathrm{s}$. The goal of the simple model is to give a clear derivation of how $T_\mathrm{s}$ depends on the parameters of the problem. Since the phase-locking arises from hydrodynamic interactions, which vanish when $a/D\rightarrow0$, we expect $T_\mathrm{s}$ to scale as some power of $D/a$ for large $D/a$.

\subsection{Dimensionless far-field equations of motion for $\theta_i$ and $\mathbf{x}_i$}
Since the balls are far apart, we expand the equations of motion in powers of $1/D$, assuming that $a\ll D$ and $R\ll D$.  Measuring length in units of $R$, time in units of $T_1$, and using force balance to eliminate $\mathbf{f}_i$ leads to the dimensionless equations of motion,
\begin{equation}
(\dot{\mathbf{ X}}+\dot{\bm{\Theta}})=-\mathsf{H}\mathbf{X}/\epsilon.\label{dimlesse4}
\end{equation}
In Eq.~(\ref{dimlesse4}), $\epsilon=M_1/(kR^2)\ll1$, $\mathbf{X}$ and $\dot{\bm{\Theta}}$ are 4$\times$1 vectors with
\begin{equation}
\mathbf{X}=\begin{pmatrix}\mathbf{x}_1\\ \mathbf{x}_2\end{pmatrix}, \quad \dot{\bm{\Theta}}=\begin{pmatrix} \dot\theta_1\hat{\bm{\theta}}_1 \\ \dot\theta_2\hat{\bm{\theta}}_2 \end{pmatrix},\label{XTheta}
\end{equation}
where $\hat{\bm{\theta}_i}=(-\sin\theta_i,\cos\theta_i)$.
The 4$\times$4 matrix $\mathsf{H}$ is the  Oseen tensor to leading order in $a/D$,
\begin{equation}
\mathsf{H}=\begin{bmatrix} \mathsf{I} & \frac{3}{4}\frac{a}{D}(\mathsf{I}+\hat{\mathbf{x}}\hat{\mathbf{x}})\\
\frac{3}{4}\frac{a}{D}(\mathsf{I}+\hat{\mathbf{x}}\hat{\mathbf{x}})& \mathsf{I} \end{bmatrix},
\end{equation}
where $\mathsf{I}$ is the 2$\times$2 identity matrix and $\hat{\mathbf{x}}\hat{\mathbf{x}}$ is the 2$\times$2 matrix with unity in the upper left-hand corner and zeros elsewhere.

Using $M_1$ as the unit for torque, the moment balance equations~(\ref{etorquebal}) take the form
\begin{eqnarray}
1+(\mathbf{x}_1/\epsilon)\cdot\hat{\theta}_1&=&0\label{mbal1}\\
1+\frac{\Delta M}{M_1}+(\mathbf{x}_2/\epsilon)\cdot\hat{\theta}_2&=&0,\label{mbal2}
\end{eqnarray}
where $\Delta M=M_2-M_1$. From these equations we conclude that  $\mathbf{x}_i$ is $\mathcal{O}(\epsilon)$. Note that the shafts have a nonzero displacement $\mathbf{x}_i$ even when the paddles are isolated.

\subsection{Far-field equations of motion for average angular speed and phase difference}
To understand phase-locking, it is not necessary to resolve the motion of the paddles on the short time scale $T_k$. In dimensionless variables, these short-scale motions are characterized by transients of the form $\exp(-t/\epsilon)$. By considering dimensionless times $t\gg\epsilon$ we may neglect these transients and treat $\epsilon\dot{\mathbf{X}}$ as small. Physically, this approximation reflects the fact that once the transients have decayed, the drag forces incurred by the small motions $\mathbf{X}$ arising from the extension of the springs are small, but not negligible, compared to the drag forces due to the rotation $\dot{\bm{\Theta}}$ of the balls about the shafts.  Therefore,  we solve Eq.~(\ref{dimlesse4}) for $\mathbf{X}$ using iteration, finding
\begin{equation}
\mathbf{X}\approx-\epsilon\mathsf{H}^{-1}\dot{\bm{\Theta}}+\epsilon^2\mathsf{H}^{-1}\frac{\mathrm{d}}{\mathrm{d}t}
\left(\mathsf{H}^{-1}\dot{\bm{\Theta}}\right).\label{Xeqnmx}
\end{equation}
In terms of $\theta_i$, we have
\begin{equation}
\mathbf{X}\approx\epsilon\begin{pmatrix}-\dot\theta_1\hat{\bm{\theta}}_1+\frac{3}{2}\frac{a}{D}\dot\theta_2\hat{\bm{\theta}}_2\\ -\dot\theta_2\hat{\bm{\theta}}_2 +\frac{3}{2}\frac{a}{D}\dot\theta_1\hat{\bm{\theta}}_1 \end{pmatrix}+\epsilon^2\begin{pmatrix} -\dot\theta_1^2\hat{\bm{\rho}}_1+\frac{3}{2}\frac{a}{D}\dot\theta_2^2\hat{\bm{\rho}}_2\\ -\dot\theta_2^2\hat{\bm{\rho}}_2 +\frac{3}{2}\frac{a}{D}\dot\theta_1^2\hat{\bm{\rho}}_1 \end{pmatrix},\label{Xexplicit}
\end{equation}
where we have only retained terms of $\mathcal{O}(a/D)$.

In Eq.~(\ref{Xexplicit}),  we have  discarded terms of the form $\ddot\theta_i$, since they are $\mathcal{O}(a^2/D^2)$. To see why, observe that for time scales longer than $T_k$, the motion is characterized by two well-separated time scales, $T_1$ and $T_\mathrm{s}$. The form of the interaction suggests that $T_\mathrm{s}\propto D/a$.
To explicitly account for the multiple scales $T_1$ and $T_\mathrm{s}$, write~\cite{Strogatz1994}
\begin{equation}
\theta_{1,2}=\omega({\tau})t\mp\Delta\theta(\tau)/2,\label{multscales}
\end{equation}
where $\tau=at/D$ describes the slowly-varying time dependence of the rotational frequency and the phase difference. Note that $\omega(\tau)$ is the average angular speed, and $\Delta\theta$ is the average phase difference. The angular speed and phase difference also have rapidly vary parts with zero average, but these are lower order in $a/D$~\cite{Strogatz1994}. Equation~(\ref{multscales}) shows that the leading term of $\ddot{\theta}_i$ is $(a/D)\omega'(\tau)$. But since the average rotation speed $\omega$ is constant in the absence of interactions, $\omega'(\tau)$ must be at least $\mathcal{O}(a/D)$. Thus, $\ddot\theta_i$ is at least $\mathcal{O}(a^2/D^2)$.

To find the governing equations for angular speed $\omega$ and phase difference $\Delta\theta$, substitute the shaft displacements $\mathbf{x}_i$ from Eq.~(\ref{Xexplicit}) into moment balance, Eqs.~(\ref{mbal1}--\ref{mbal2}). Finally, average the resulting equations over a period, treating the slowly-varying variables $\omega$ and $\Delta\theta$ as constants under the average. We find that the average dimensionless speed is given by
\begin{equation}
\omega = 1+\frac{\Delta M}{M_1}+\frac{9}{8}\frac{a}{D}\cos\Delta\theta.\label{asymomega}
\end{equation}
The interacting paddles turn faster than they would in isolation. This result is in contrast with our paddle experiments, where we saw in \S\ref{expsec} that the asymmetric paddles rotated more slowly compared to an isolated paddle. It is too much to demand that our far-field theory captures every aspect of the paddle experiments, since the paddles are close to each other in the experiment and the theory is valid when they are far apart.

The dimensionless phase difference obeys
\begin{equation}
\frac{\mathrm d  \Delta\theta}{\mathrm d t}=-\frac{9}{2}\epsilon\frac{a}{D}\sin\Delta\theta+\frac{\
\Delta M}{M_1}.\label{asymtheory}
\end{equation}
These results~(\ref{asymomega}--\ref{asymtheory}) are equivalent to the results of reference~\cite{niedermayer08}.
For equal driving torques, $\Delta M=0$, Eq.~(\ref{asymtheory})
shows that the paddles synchronize to $\Delta\theta=0$, independent of the initial value of $\Delta\theta$, in (dimensional) time  $T_\mathrm{s}\sim (D/a)(k R^2/M_1)T_1$, or
\begin{equation}
T_\mathrm{s}\sim\frac{D}{a}\frac{kR^2}{M_1}\frac{6\pi\eta aR^2}{M_1}.
\end{equation}
When $M_1\neq M_2$, the paddles phase-lock with a nonzero phase difference, which increases to $\pi/2$ in the steady state as the torque difference increases to the critical value given by
$\Delta M/M_1=(9/4)(a/D)M_1/(kR^2)$. Note that the factor of $a/D$ and the smallness of $\epsilon=M_1/(kR^2)$ mean that $M_2$ must be very close to $M_1$ for the  phase difference $\Delta\theta$ to have a fixed point. Thus, in the derivation of Eqs.~(\ref{asymomega}--\ref{asymtheory}) we  considered $\Delta M/M_1$ and $T_1\Delta\dot{\theta}$ to be $\mathcal{O}(\epsilon a/D)$.

This simple theory predicts that $T_\mathrm{s}/T_1$ varies inversely with torque, whereas the experiments show that $T_\mathrm{s}/T_1$ depends at most weakly on torque (Fig.~\ref{fig:scaletor}). Again, the resolution of this discrepancy is that the simple theory is valid in the far-field limit with  $D\gg a$, whereas the experiments are carried out in the near-field regime where $T_\mathrm{s}/T_1$ is independent of torque.

\subsection{Physical explanation for phase locking}

\begin{figure}
\includegraphics[width=3.4in]{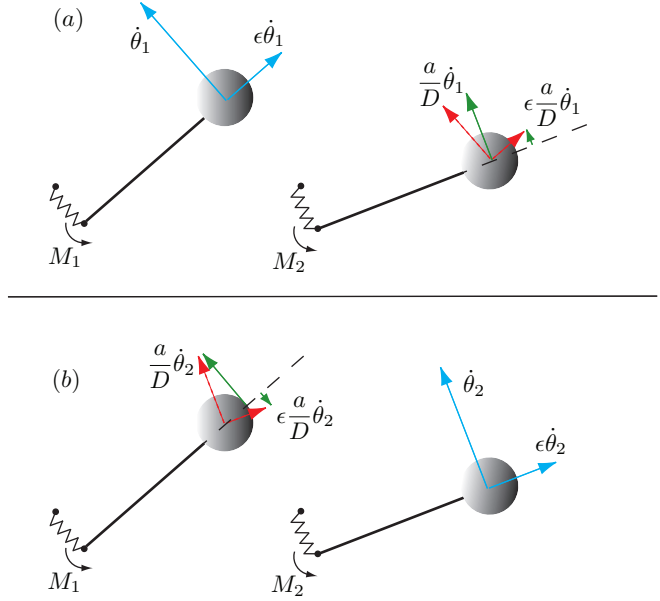}
\caption{(Color online.) Physical explanation for synchronization. The figure is not to scale and the paddles have been moved artificially close. (a) On the left we show the components of velocity of the first ball (blue arrows). On the right, we show the components of the drag induced on the second ball by the motion of the first ball (red dotted arrows), and the components that contribute to the hydrodynamic torque on the second paddle (green arrows). (b) The same situation as (a), but showing the velocity components of the second ball and the induced forces on the first.  The difference in the hydrodynamic torques tends to make $\Delta\theta=0$.}
\label{fig:physicalexplanation}
\end{figure}

Each of the terms of Eq.~(\ref{Xexplicit}) has a simple interpretation. First consider the limit of an isolated paddle, $a/D=0$. To leading order in $\epsilon$,  the ball on the end of the rod undergoes circular motion. This motion leads to a drag in the $-\hat{\theta}_i$ direction, which stretches the spring along $-\hat{\theta}_i$, which in turn leads to an $\mathcal{O}(\epsilon)$ component of the ball's velocity parallel to the rod, along the $\hat{\rho}_i$ direction (see the left ball in Fig.~\ref{fig:physicalexplanation}a). In our dimensionless units, the ball exerts an $\mathcal{O}(1)$ force on the liquid in the $\hat{\theta}_i$ direction, and an $\mathcal{O}(\epsilon)$ force on the liquid in the $\hat{\rho}_i$ direction.  To get the displacement $\mathbf{X}$, we multiply these forces by $\epsilon$, and thus get the $\mathcal{O}[(a/D)^0]$ terms of Eq.~(\ref{Xexplicit}). Now consider the hydrodynamic interactions. For a given paddle, each of the forces just described induces a Stokeslet flow, falling off inversely with distance, and leading to the $\mathcal{O}(a/D)$ terms in Eq.~(\ref{Xexplicit}).

The drag forces on each paddle induced by the motion of the other are shown in Fig.~\ref{fig:physicalexplanation}. From this figure we can see why the paddles synchronize. Suppose that the second paddle slightly lags the first. Since the spring is flexible, the ball of the paddle on the left has a velocity component of $\epsilon\dot{\theta}_1$ along the rod  as well as the component $\dot{\theta}_1$ perpendicular to the rod (blue arrows, Fig.~\ref{fig:physicalexplanation}a, left). This motion induces drag forces on the ball on the right (red dotted arrows, Fig.~\ref{fig:physicalexplanation}a, right), which in our dimensionless units are down by a factor of $a/D$ from the velocities. The components of these forces perpendicular to the rod (green arrows, Fig.~\ref{fig:physicalexplanation}a, right) contribute to the hydrodynamic torque on the paddle. Likewise the motion of the paddle on the right (blue arrows, Fig.~\ref{fig:physicalexplanation}b, left) induces forces that lead to hydrodynamic torques on the left paddle. The phase difference $\Delta\theta$ is governed by the difference of the torques, which for small $\Delta\theta$ is given by the difference of the small (green) arrows in Figs.~\ref{fig:physicalexplanation}a and b. The torque difference makes $\Delta\theta=0$ a stable fixed point (for $\Delta M=0$).

\subsection{Power dissipation}
We may readily examine the question of power dissipation using our simple model. First note that for fixed driving torques, the power dissipated decreases when the hydrodynamic resistance of the paddles increases. Therefore,
when $\Delta\theta=0$, the drag is minimized and the dissipation rate is maximized. As $\Delta M$ increases, the increase in $\Delta\theta$ leads to greater resistance and therefore lower dissipation rate. To leading order in $\epsilon$, we may use Eq.~(\ref{asymomega}) to show that the dimensionless power averaged over one period, $P=M_1\dot\theta_1+M_2\dot\theta_2$,  takes the form
\begin{equation}
\frac{P}{M_1}=2+\frac{9}{4}\frac{a}{D}\cos\Delta\theta.
\end{equation}
In general, the phase difference chosen by the system does not minimize the power dissipated. The same conclusion has been reached for the hydrodynamic phase-locking of nearby swimming sheets~\cite{ElfringLauga2009}.

\section{Simple Model for Symmetric Paddles}
\label{symsec}
\begin{figure}
\includegraphics[width=3.5in]{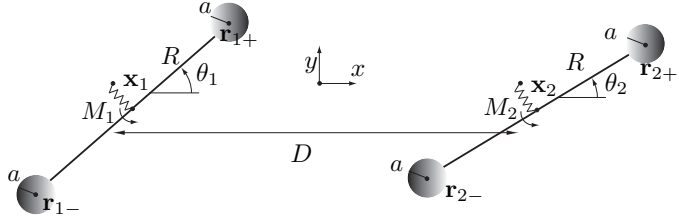}
\caption{Model for symmetric paddles. The figure is not to scale; note that $a\ll R\ll D$.}
\label{fig:twoball}
\end{figure}
\subsection{Oseen model and nondimensionalization}
To understand why the symmetric paddles lock phases with $\Delta\theta=\pi/2$ when $\Delta M=0$, we model the paddles as dumbells (Fig.~\ref{fig:twoball}). Each dumbell consists of two balls connected by a rod that does not disturb the fluid as it moves. The midpoint of each rod is attached to a fixed point by a stiff spring, and the balls at the ends of the rods have positions
\begin{eqnarray}
\mathbf{r}_{1\pm}&=&-(D/2)\hat{\mathbf{x}}+\mathbf{x}_1\pm R\hat{\bm{\rho}}_1\\
\mathbf{r}_{2\pm}&=&(D/2)\hat{\mathbf{x}}+\mathbf{x}_2\pm R\hat{\bm{\rho}}_2,
\end{eqnarray}
where $\mathbf{x}_i$ is the displacement of the midpoint of the $i$th rod from the corresponding fixed point.
Denoting by $\mathbf{f}_{i\pm}$ the forces that the balls on the $i$th dumbell exert on the fluid, the balance of forces on each dumbell implies
\begin{equation}
-\mathbf{f}_{i+}-\mathbf{f}_{i-}-k\mathbf{x}_i=\mathbf{0}\label{Fipm},
\end{equation}
and the balance of torques implies
\begin{equation}
M_i+\hat{\mathbf{z}}\cdot\left(R\hat{\bm{\rho}}_i\right)\times\left(-\mathbf{f}_{i+}\right)
+\hat{\mathbf{z}}\cdot\left(-R\hat{\bm{\rho}}_i\right)\times\left(-\mathbf{f}_{i-}\right)=0.\label{mbalSymm}
\end{equation}
Assuming all balls are far apart, we again use the Oseen model, Eq.~(\ref{Oseenmodel}), this time extended to the four balls labeled $\alpha=1-,1+,2-,2+$: 
\begin{equation}
\mathbf{v}_\alpha=\frac{\mathbf{f}_\alpha}{6\pi\eta a}+\frac{1}{8\pi\eta}\sum_{\beta\neq \alpha}\left[\frac{\mathbf{f}_\beta}{|\mathbf{r}_{\alpha\beta}|}+\frac{(\mathbf{f}_\beta\cdot\mathbf{r}_{\alpha\beta})\mathbf{r}_{\alpha\beta}}{|\mathbf{r}_{\alpha\beta}|^3}\right],\label{Oseenmodel_general}
\end{equation}
where $\mathbf{r}_{\alpha\beta}=\mathbf{r}_\alpha-\mathbf{r}_\beta$. This is valid when $R\gg a$ and $D\gg a$, but we will also assume $D\gg R$.

The $\theta_i\mapsto\theta_i+\pi$ symmetry of the dumbells makes the hydrodynamic interaction between the dumbells more subtle than the asymmetric case. First observe that the spring of an isolated rotating dumbell does not stretch  since the net hydrodynamic force on the balls vanishes. Thus, $\mathbf{x}_i=\mathbf{0}$ when $D\rightarrow\infty$. However, for finite $D/R$, the flow induced by the rotation of one dumbell causes the spring of the other dumbell to stretch. To estimate the amount of stretch, consider the flow induced by dumbell 1 at dumbell 2. The far-field flow is an asymmetric force dipole, also known as a rotlet, falling off inversely with the square of distance~\cite{russel_etal1989}.  Thus, the flow $v_{21}$ induced at dumbell 2 is approximately $v_{21}\sim {f_1}R/(\eta D^2)$, leading to drag on dumbell 2 of about $\eta a v_{21}\sim a f_1 R/D^2\sim aM_1/D^2$.
This drag causes the spring of dumbell to stretch, with a displacement
\begin{equation}
\frac{|\mathbf{x}_2|}{R}\sim\frac{a}{R}\frac{R^2}{D^2}\frac{M_1}{k R^2}.
\end{equation}

As in the previous section, it is convenient to measure length in units of $R$ and time in units of $T_1=6\pi\eta R^2a/M_1$. Thus, the dimensionless displacement is $|x_2|\sim\epsilon a/D^2$.

\subsection{Far-field equations of motion}

For each spring of a pair of rotating dumbells, the leading order stretch of the spring is second order in $D^{-1}$. However, to derive equations describing phase-locking of symmetric dumbells, we will see that we must expand the displacements $\mathbf{x}_i$ to $\mathcal{O}(D^{-3})$. These third order displacements arise from reflections of the dipole force. For example, the $\mathcal{O}(D^{-2})$ deflection of spring 2 from the dipole originating  at dumbell 1 induces a point force at dumbell 2. This point force  causes an $\mathcal{O}(D^{-3})$ displacement at dumbell 1.
As we will see, to find $\mathbf{x}_i$ to  $\mathcal{O}(D^{-3})$ we need only expand the Oseen tensor to second order in $D^{-1}$. Thus, to second order in $D^{-1}$, Eq.~(\ref{Oseenmodel_general}) becomes
\begin{equation}
\begin{pmatrix} \dot{\mathbf{x}}_1+\dot\theta_1\hat{\bm{\theta}}_1 \\ \dot{\mathbf{x}}_1-\dot\theta_1\hat{\bm{\theta}}_1 \\ \dot{\mathbf{x}}_2+\dot\theta_2\hat{\bm{\theta}}_2  \\ \dot{\mathbf{x}}_2-\dot\theta_2\hat{\bm{\theta}}_2 \end{pmatrix}=\mathsf{H}\begin{pmatrix}\mathbf{f}_{1+}\\ \mathbf{f}_{1-}\\ \mathbf{f}_{2+}\\ \mathbf{f}_{2-}\end{pmatrix},\label{H8x8}
\end{equation}
where $\mathsf{H}$ is the  $8\times8$ Oseen tensor evaluated at $\mathbf{x}_1=\mathbf{x}_2=\mathbf{0}$ and expanded to $\mathcal{O}(D^{-2})$.

Before describing $\mathsf{H}$, it is convenient to re-express Eq.~(\ref{H8x8}) in terms of the sums and differences of forces on each paddle, $\mathbf{f}_i=\mathbf{f}_{i+}+\mathbf{f}_{i-}$ and $\Delta\mathbf{f}_i=\mathbf{f}_{i+}-\mathbf{f}_{i-}$, respectively. Note that in terms of these variables, the dimensionless force-balance equation~(\ref{Fipm}) becomes
\begin{equation}
-\mathbf{f}_i-\mathbf{x}_i/\epsilon=\mathbf{0},\label{dimlessFipm}
\end{equation}
and the dimensionless moment-balance equations become
\begin{eqnarray}
1-\Delta\mathbf{f}_1\cdot\hat{\theta}_1&=&0\label{dimlessmbal1}\\
1+\frac{\Delta M}{M_1}-\Delta\mathbf{f}_2\cdot\hat{\theta}_2&=&0,
\label{dimlessmbal2}
\end{eqnarray}
where $M_1$ is the unit for torque.
Returning to Eq.~(\ref{H8x8}), we add and subtract the appropriate rows of Eq.~(\ref{H8x8}) and rearrange to find
\begin{eqnarray}
\dot{\mathbf{X}}&=&\frac{1}{\epsilon}\mathsf{A}\mathbf{X}+\mathsf{B}\Delta\mathbf{F}\label{XAB}\\
\dot{\bm{\Theta}}&=&\frac{1}{\epsilon}\mathsf{C}\mathbf{X}+\mathsf{D}\Delta\mathbf{F}.\label{ThAB}
\end{eqnarray}
where $\mathbf{X}$ and $\dot{\bm{\Theta}}$ are defined as before in Eq.~(\ref{XTheta});
the $4\times4$ matrices $\mathsf{A}$, $\mathsf{B}$, $\mathsf{C}$, and $\mathsf{D}$ are given in the appendix; and
\begin{equation}
\Delta\mathbf{F}=\begin{pmatrix}\Delta\mathbf{f}_1\\\Delta\mathbf{f}_2\end{pmatrix}.
\end{equation}

Expanding in powers of $D^{-1}$, we find (see Appendix)
\begin{eqnarray}
\mathsf{A}&=&\mathsf{A}^{(0)}+D^{-1}\mathsf{A}^{(1)}+\mathcal{O}(D^{-3})\\
\mathsf{B}&=&D^{-2}\mathsf{B}^{(2)}+\mathcal{O}(D^{-3})\\
\mathsf{C}&=&D^{-2}\mathsf{C}^{(2)}+\mathcal{O}(D^{-3})\\
\mathsf{D}&=&\mathsf{D}^{(0)}+\mathcal{O}(D^{-3}).
\end{eqnarray}
Likewise, we expand $\mathbf{X}$ and $\Delta\mathbf{F}$ in powers of $D^{-1}$:
\begin{eqnarray}
\mathbf{X}&=&\mathbf{X}^{(0)}+D^{-1}\mathbf{X}^{(1)}+D^{-2}\mathbf{X}^{(2)}+\cdots\\
\Delta{\mathbf{F}}&=&\Delta{\mathbf{F}}^{(0)}+D^{-1}\Delta{\mathbf{F}}^{(1)}+D^{-2}\Delta{\mathbf{F}}^{(2)}+\cdots.
\end{eqnarray}
Since $\mathsf{B}$ and $\mathsf{C}$ are $\mathcal{O}(D^{-2})$ at leading order, our order of expansion is sufficient for determining $\mathbf{X}$ to $\mathcal{O}(D^{-3})$ and $\Delta\mathbf{F}$ to $\mathcal{O}(D^{-5})$.  At zeroth order, we find $\mathbf{X}^{(0)}=\mathbf{0}$, as expected, and  $\Delta\mathbf{F}^{(0)}=[\mathsf{D}^{(0)}]^{-1}\dot{\bm{\Theta}}$, with
\begin{equation}
\Delta\mathbf{f}_i^{(0)}= \frac{2\dot{\theta}_i\hat{\bm{\theta}}_i}
{1-3a/8}.\label{Dfi0}
\end{equation}
Substituting $\Delta\mathbf{f}^{(0)}$ into the moment balance equations~(\ref{dimlessmbal1}--\ref{dimlessmbal2}), taking their sum and difference, using Eq.~(\ref{multscales}) to eliminate $\theta_1$ and $\theta_2$ in favor of $\omega$ and $\Delta\theta$, and integrating over one period yields the dimensionless average speed and phase difference,
\begin{eqnarray}
\omega^{(0)}&=&1/2\left(1-3a/8\right)\label{wDth01}\\
\Delta\dot{\theta}^{(0)}&=&\left(1-3a/8\right)\Delta M/M_1.\label{wDth02}
\end{eqnarray}
Since $\omega^{(0)}$ and $\Delta\dot{\theta}^{(0)}$ are independent of $D$, they are the average speed and phase-difference, respectively, for non-interacting dumbells. There is no phase-locking if there is no interaction, and the phase difference increases in proportion to the difference in driving torques, $\Delta M$.

Note that the factors of $3a/8$ in Eq.~(\ref{Dfi0}) are due to the interaction between the two balls of a given dumbell:
one ball induces a disturbance flow of magnitude $(6\pi\eta a\dot{\theta})/(8\pi\eta 2)=3a\dot{\theta}/8$ at the other ball. This disturbance flow hinders the motion of the other ball.

The leading order displacements of the shafts are given by
\begin{equation}
\dot{\mathbf{X}}^{(2)}=\frac{1}{\epsilon}\mathsf{A}^{(0)}\mathbf{X}^{(2)}+\mathsf{B}^{(2)}\Delta\mathbf{F}^{(0)}.
\end{equation}
As in the case of the asymmetric paddles, this equation is readily solved to $\mathcal{O}(\epsilon^2)$; however, the full expression is so cumbersome that we only report the result to leading order in $\epsilon$ and $a$ in the appendix. The next order contribution to the force difference is given by
\begin{equation}
0=\frac{1}{\epsilon}\mathsf{C}^{(2)}\mathbf{X}^{(2)}+\mathsf{D}^{(0)}\Delta\mathbf{F}^{(4)}.
\end{equation}
Again, the full expression for $\Delta\mathbf{F}^{(4)}$ is so cumbersome that we only report the leading order terms in the appendix. Using $\Delta\mathbf{F}^{(4)}$ in the difference of the moment equations and averaging yields terms proportional to $\Delta\dot{\theta}$, which do not lead to phase-locking.
The average of the sum of the moment equations leads to a decrease in the average rotation speed, which together with Eqn.~(\ref{wDth01}) yields
\begin{equation}
\omega =\frac{1}{2}-\frac{3a}{16}-\frac{153}{16}\frac{a^2}{D^4}.\label{omegaasymm}
\end{equation}
The interacting paddle turn more slowly than they would in isolation.

The third-order displacement of the shafts is determined by
\begin{equation}
\dot{\mathbf{X}}^{(3)}=\frac{1}{\epsilon}\mathsf{A}^{(1)}\mathbf{X}^{(2)}+\frac{1}{\epsilon}\mathsf{A}^{(0)}\mathbf{X}^{(3)}.
\end{equation}
Solving for $\mathbf{X}^{(3)}$ (see Appendix for leading terms), and substituting into
\begin{equation}
0=\frac{1}{\epsilon}\mathsf{C}^{(2)}\mathbf{X}^{(3)}+\mathsf{D}^{(0)}\Delta\mathbf{F}^{(5)}
\end{equation}
yields $\Delta\mathbf{F}^{(5)}$ (see Appendix for leading terms), which has terms that lead to phase-locking. Using moment balance Eqs.~(\ref{dimlessmbal1}--\ref{dimlessmbal2}) and averaging, together with the leading order result (\ref{Dfi0}), yields
\begin{equation}
\Delta\dot\theta =\frac{1}{2}\Delta M+\frac{243}{8}\epsilon\frac{a^3}{D^5}\sin2\Delta\theta\label{synchsym}
\end{equation}

Equation~(\ref{synchsym}) is the main result of this section. The (dimensional) synchronization time for the symmetric paddles scales as
\begin{equation}
T_\mathrm{s}\sim\frac{D^5}{a^3 R^2}\frac{kR^2}{M_1}\frac{6\pi\eta aR^2}{M_1}.
\end{equation}
When $\Delta M=0$, Eq.~(\ref{synchsym}) has a stable fixed point at $\Delta\theta=\pi/2$, in accord with our experiments and the more accurate regularized stokeslet simulation of \S\ref{numersec}. As in the case of the asymmetric paddles, the torque difference must be small for phase-locking to occur. The critical torque difference, above which phase-locking cannot occur, is $\Delta M_{\mathrm{crit}}=(243/4)\epsilon a^3/D^5$. Note that the average phase-difference in the phase-locked state depends on $\Delta M$. Note also that the time for phase-locking depends more strongly on separation for the symmetric paddles compared to the asymmetric paddles. It is not easy to give a simple physical picture for why the paddle separation $D$ enters the synchronization time with a  fifth power. We simply note two effects: (1) the flow induced by the force dipole of one paddle reflects off the other paddle, and then again off the first paddle, leading to four powers of $D^{-1}$, and (2) the torque exerted by a flow on the paddle arises from the difference in the flow at the two ends of the paddle, leading to another factor of $D^{-1}$. Our case is reminiscent of the fifth power that appears in the reorientation of oscillating  dumbells~\cite{alexander08}.  Although  our experiments were not carried out in the far-field regime, we found that
the synchronization time depends more strongly on separation in the symmetric case compared to the asymmetric case (Fig.~\ref{fig:gap_syn}). Finally, we note that to leading order in $D^{-1}$, the power dissipated in the synchronized state is independent of $\Delta\theta$, since the average rotation speed $\omega$ in the synchronized state is independent of $\Delta\theta$ (Eqn.~\ref{omegaasymm}).

\section{Conclusion}

To summarize, we have presented perhaps the simplest experimental realization of the phenomenon of hydrodynamic synchronization at low Reynolds number. The requirements for synchronization are subtle: the system must have a slight flexibility to allow small shifts in the positions of the paddles. Since this flexibility is generic, we expect that conditions allowing hydrodynamic synchronization will commonly arise in a wide range of systems at low Reynolds number. On the other hand, our work indicates that hydrodynamic synchronization is not robust, since it requires that the driving moments be fine-tuned to be close to each other.

This work was supported in part by National Science Foundation Grants Nos. CTS-0828239 (KSB), NIRT-0404031 (TRP), DMS-0615919 (TRP), and CBET-0854108 (TRP \& KSB). We thank D. Bartolo, R. Cortez,  J. Elgeti, R. Goldstein,  I. Tuval, and especially L. Setayeshgar and  H. Fu for helpful conversations and comments. TRP thanks the Aspen Center for Physics where some of this work was completed.

\appendix\section{Oseen tensor for symmetric paddles}

For $\mathsf{H}$ evaluated at $\mathbf{x}_i=\mathbf{0}$ and the matrices $\mathsf{A}$, $\mathsf{B}$, $\mathsf{C}$, and $\mathsf{D}$ defined in Eqs.~(\ref{XAB}--\ref{ThAB}), expanding in powers of $D^{-1}$ yields
\begin{widetext}
\begin{equation}
\mathsf{A}=\begin{pmatrix}-\frac{1}{2}-\frac{9a}{32}-\frac{3}{32}a\cos(2\theta_1)&-\frac{3}{32}a\sin(2\theta_1)&-\frac{3a}{2D}& 0\\
-\frac{3}{32}a\sin(2\theta_1) & -\frac{1}{2}-\frac{9a}{32}+\frac{3}{32}a\cos(2\theta_1) & 0 & -\frac{3a}{4D}\\
-\frac{3a}{2D} & 0 &-\frac{1}{2}-\frac{9a}{32}-\frac{3}{32}a\cos(2\theta_2)&  -\frac{3}{32}a\sin(2\theta_2)\\
0 & -\frac{3a}{4D} & -\frac{3}{32}a\sin(2\theta_2) & -\frac{1}{2}-\frac{9a}{32}+\frac{3}{32}a\cos(2\theta_2)
\end{pmatrix}
\end{equation}
and
\begin{equation}
\mathsf{D}=
\begin{pmatrix}-\frac{1}{2}-\frac{9a}{32}-\frac{3}{32}a\cos(2\theta_1)&-\frac{3}{32}a\sin(2\theta_1)&0& 0\\
-\frac{3}{32}a\sin(2\theta_1) & \frac{1}{2}-\frac{9a}{32}+\frac{3}{32}a\cos(2\theta_1) & 0 &0\\
0 & 0 &-\frac{1}{2}-\frac{9a}{32}-\frac{3}{32}a\cos(2\theta_2)&  -\frac{3}{32}a\sin(2\theta_2)\\
0 & 0 & -\frac{3}{32}a\sin(2\theta_2) & \frac{1}{2}-\frac{9a}{32}+\frac{3}{32}a\cos(2\theta_2)
\end{pmatrix}
\end{equation}
\end{widetext}
for the blocks on the diagonal of the Oseen tensor. For the blocks off the diagonal, we have
\begin{eqnarray}
\mathsf{B}&=&\begin{pmatrix} 0 & 0 & -\frac{3a^2}{2D^2}\cos\theta_2&\frac{3a^2}{4D^2}\sin\theta_2\\
0 & 0 & \frac{3a^2}{4D^2}\sin\theta_2 & -\frac{3a^2}{2D^2}\cos\theta_2\\
\frac{3a^2}{2D^2}\cos\theta_1 & -\frac{3a^2}{4D^2}\sin\theta_1& 0 & 0\\
-\frac{3a^2}{4D^2}\sin\theta_1 & \frac{3a^2}{2D^2}\cos\theta_1 & 0 & 0
\end{pmatrix}\\
\mathsf{C}&=&
\begin{pmatrix}
0 & 0& -\frac{3a}{2D^2}\cos\theta_1&\frac{3a}{4D^2}\sin\theta_1\\
0 & 0 & \frac{3a}{4D^2}\sin\theta_1 & -\frac{3a}{4D^2}\cos\theta_1\\
\frac{3a}{2D^2}\cos\theta_2& -\frac{3a}{4D^2}\sin\theta_2& 0 & 0 \\
-\frac{3a}{4D^2}\sin\theta_2&\frac{3a}{4D^2}\cos\theta_2&0 &0
\end{pmatrix}.
\end{eqnarray}

The second order spring deflection to leading order  in $\epsilon$ and $a$ is
\begin{equation}
\mathbf{X}^{(2)}=a\epsilon\begin{pmatrix}(9/2) \dot{\theta}_2\sin(2\theta_2)\\
-3 \dot{\theta}_2\\
-(9/2) \dot{\theta}_1\sin(2\theta_1)\\
3 \dot{\theta}_1\end{pmatrix}+\mathcal{O}(\epsilon a^2) .
\end{equation}
The fourth order force difference to leading order in $a$ and $\epsilon$ is
\begin{equation}
\Delta\mathbf{F}^{(4)}\approx\frac{9}{4}a^2
\begin{pmatrix}
-\dot{\theta}_1\left[5\sin\theta_1+3\sin(3\theta_1)\right]\\
2\dot{\theta}_1\cos\theta_1\left(1+3\sin^2\theta_1\right)\\
-\dot{\theta}_2\left[5\sin\theta_2+3\sin(3\theta_2)\right]\\
2\dot{\theta}_2\cos\theta_2\left(1+3\sin^2\theta_2\right)
\end{pmatrix}.
\end{equation}
The third order spring deflection, leading order in $\epsilon$ and $a$,
\begin{equation}
\mathbf{X}^{(3)}=\frac{9}{2}a^2\epsilon\begin{pmatrix} 3\dot{\theta}_1\sin(2\theta_1)\\
-\dot{\theta}_1\\
-3\dot{\theta}_2\sin(2\theta_2)\\
\dot{\theta}_2\end{pmatrix}.
\end{equation}
Finally, again to leading order in $\epsilon$ and $a$,
\begin{equation}
\Delta\mathbf{F}^{(5)}=\frac{27}{4}a^3
\begin{pmatrix}
-\dot{\theta}_2\left[\sin\theta_1+6\cos\theta_1\sin(2\theta_2)\right]\\
\dot{\theta}_2\left[\cos\theta_1+3\sin\theta_1\sin(2\theta_2)\right]\\
-\dot{\theta}_1\left[\sin\theta_2+6\cos\theta_2\sin(2\theta_1)\right]\\
\dot{\theta}_1\left[\cos\theta_2+3\sin\theta_2\sin(2\theta_1)\right]
\end{pmatrix}.
\end{equation}

\bibliographystyle{apsrev}

\bibliography{synrefs}

\begin{thebibliography}{27}
\expandafter\ifx\csname natexlab\endcsname\relax\def\natexlab#1{#1}\fi
\expandafter\ifx\csname bibnamefont\endcsname\relax
  \def\bibnamefont#1{#1}\fi
\expandafter\ifx\csname bibfnamefont\endcsname\relax
  \def\bibfnamefont#1{#1}\fi
\expandafter\ifx\csname citenamefont\endcsname\relax
  \def\citenamefont#1{#1}\fi
\expandafter\ifx\csname url\endcsname\relax
  \def\url#1{\texttt{#1}}\fi
\expandafter\ifx\csname urlprefix\endcsname\relax\def\urlprefix{URL }\fi
\providecommand{\bibinfo}[2]{#2}
\providecommand{\eprint}[2][]{\url{#2}}

\bibitem[{\citenamefont{Machemer}(1972)}]{machemer1972}
\bibinfo{author}{\bibfnamefont{H.}~\bibnamefont{Machemer}},
  \bibinfo{journal}{J. Exp. Biol.} \textbf{\bibinfo{volume}{{\bf 57}}},
  \bibinfo{pages}{239} (\bibinfo{year}{1972}).

\bibitem[{\citenamefont{Polin et~al.}(2009)\citenamefont{Polin, Tuval,
  Drescher, Gollub, and Goldstein}}]{PolinTuvalDrescherGollubGoldstein2009}
\bibinfo{author}{\bibfnamefont{M.}~\bibnamefont{Polin}},
  \bibinfo{author}{\bibfnamefont{I.}~\bibnamefont{Tuval}},
  \bibinfo{author}{\bibfnamefont{K.}~\bibnamefont{Drescher}},
  \bibinfo{author}{\bibfnamefont{J.~P.} \bibnamefont{Gollub}},
  \bibnamefont{and} \bibinfo{author}{\bibfnamefont{R.~E.}
  \bibnamefont{Goldstein}}, \bibinfo{journal}{Science}
  \textbf{\bibinfo{volume}{325}}, \bibinfo{pages}{487} (\bibinfo{year}{2009}).

\bibitem[{\citenamefont{Goldstein et~al.}(2009)\citenamefont{Goldstein, Polin,
  and Tuval}}]{GoldsteinPolinTuval2009}
\bibinfo{author}{\bibfnamefont{R.~E.} \bibnamefont{Goldstein}},
  \bibinfo{author}{\bibfnamefont{M.}~\bibnamefont{Polin}}, \bibnamefont{and}
  \bibinfo{author}{\bibfnamefont{I.}~\bibnamefont{Tuval}},
  \bibinfo{journal}{Phys. Rev. Lett.} \textbf{\bibinfo{volume}{103}},
  \bibinfo{pages}{168103} (\bibinfo{year}{2009}).

\bibitem[{\citenamefont{Riedel et~al.}(2005)\citenamefont{Riedel, Kruse, and
  Howard}}]{reidel}
\bibinfo{author}{\bibfnamefont{I.~H.} \bibnamefont{Riedel}},
  \bibinfo{author}{\bibfnamefont{K.}~\bibnamefont{Kruse}}, \bibnamefont{and}
  \bibinfo{author}{\bibfnamefont{J.}~\bibnamefont{Howard}},
  \bibinfo{journal}{Science} \textbf{\bibinfo{volume}{309}},
  \bibinfo{pages}{300} (\bibinfo{year}{2005}).

\bibitem[{\citenamefont{Nonaka et~al.}(1998)\citenamefont{Nonaka, Tanaka,
  Takeda, Harada, Kanai, Kido, and Hirokawa}}]{nonaka_etal1998}
\bibinfo{author}{\bibfnamefont{S.}~\bibnamefont{Nonaka}},
  \bibinfo{author}{\bibfnamefont{Y.}~\bibnamefont{Tanaka}},
  \bibinfo{author}{\bibfnamefont{S.}~\bibnamefont{Takeda}},
  \bibinfo{author}{\bibfnamefont{A.}~\bibnamefont{Harada}},
  \bibinfo{author}{\bibfnamefont{Y.}~\bibnamefont{Kanai}},
  \bibinfo{author}{\bibfnamefont{M.}~\bibnamefont{Kido}}, \bibnamefont{and}
  \bibinfo{author}{\bibfnamefont{N.}~\bibnamefont{Hirokawa}},
  \bibinfo{journal}{Cell} \textbf{\bibinfo{volume}{{\bf 95}}},
  \bibinfo{pages}{829} (\bibinfo{year}{1998}).

\bibitem[{\citenamefont{Suarez and Pacey}(2006)}]{suarez06}
\bibinfo{author}{\bibfnamefont{S.~S.} \bibnamefont{Suarez}} \bibnamefont{and}
  \bibinfo{author}{\bibfnamefont{A.~A.} \bibnamefont{Pacey}},
  \bibinfo{journal}{Human Reprod. Update} \textbf{\bibinfo{volume}{12}},
  \bibinfo{pages}{23} (\bibinfo{year}{2006}).

\bibitem[{\citenamefont{Pikovsky et~al.}(2001)\citenamefont{Pikovsky,
  Rosenblum, and Kurths}}]{PikovskyRosenbluKurths2001}
\bibinfo{author}{\bibfnamefont{A.}~\bibnamefont{Pikovsky}},
  \bibinfo{author}{\bibfnamefont{M.}~\bibnamefont{Rosenblum}},
  \bibnamefont{and} \bibinfo{author}{\bibfnamefont{J.}~\bibnamefont{Kurths}},
  \emph{\bibinfo{title}{Synchronization. {A} universal concept in nonlinear
  science}} (\bibinfo{publisher}{Cambridge University Press},
  \bibinfo{address}{Cambridge}, \bibinfo{year}{2001}).

\bibitem[{\citenamefont{Gray}(1928)}]{gray1928}
\bibinfo{author}{\bibfnamefont{J.}~\bibnamefont{Gray}},
  \emph{\bibinfo{title}{Ciliary movement}} (\bibinfo{publisher}{Cambridge
  University Press}, \bibinfo{address}{Cambridge, U.K.}, \bibinfo{year}{1928}).

\bibitem[{\citenamefont{{Sleigh, ed.}}(1974)}]{Sleigh1974}
\bibinfo{author}{\bibfnamefont{M.~A.} \bibnamefont{{Sleigh, ed.}}},
  \emph{\bibinfo{title}{Cilia and flagella}} (\bibinfo{publisher}{Academic
  Press}, \bibinfo{address}{London}, \bibinfo{year}{1974}).

\bibitem[{\citenamefont{Fauci}(1990)}]{fauci90}
\bibinfo{author}{\bibfnamefont{L.~J.} \bibnamefont{Fauci}},
  \bibinfo{journal}{J. Comp. Phys.} \textbf{\bibinfo{volume}{86}},
  \bibinfo{pages}{294} (\bibinfo{year}{1990}).

\bibitem[{\citenamefont{Gueron et~al.}(1997)\citenamefont{Gueron,
  Levit-Gurevich, Liron, and Blum}}]{gueron97}
\bibinfo{author}{\bibfnamefont{S.}~\bibnamefont{Gueron}},
  \bibinfo{author}{\bibfnamefont{K.}~\bibnamefont{Levit-Gurevich}},
  \bibinfo{author}{\bibfnamefont{N.}~\bibnamefont{Liron}}, \bibnamefont{and}
  \bibinfo{author}{\bibfnamefont{J.~J.} \bibnamefont{Blum}},
  \bibinfo{journal}{Proc. Natl. Acad. Sci. USA} \textbf{\bibinfo{volume}{94}},
  \bibinfo{pages}{6001} (\bibinfo{year}{1997}).

\bibitem[{\citenamefont{Lagomarsino et~al.}(2003)\citenamefont{Lagomarsino,
  Jona, and Bassetti}}]{lagomarsino03}
\bibinfo{author}{\bibfnamefont{M.~C.} \bibnamefont{Lagomarsino}},
  \bibinfo{author}{\bibfnamefont{P.}~\bibnamefont{Jona}}, \bibnamefont{and}
  \bibinfo{author}{\bibfnamefont{B.}~\bibnamefont{Bassetti}},
  \bibinfo{journal}{Phys. Rev. E} \textbf{\bibinfo{volume}{68}},
  \bibinfo{pages}{021908} (\bibinfo{year}{2003}).

\bibitem[{\citenamefont{Lagomarsino et~al.}(2002)\citenamefont{Lagomarsino,
  Bassetti, and Jona}}]{lagomarsino02}
\bibinfo{author}{\bibfnamefont{M.~C.} \bibnamefont{Lagomarsino}},
  \bibinfo{author}{\bibfnamefont{B.}~\bibnamefont{Bassetti}}, \bibnamefont{and}
  \bibinfo{author}{\bibfnamefont{P.}~\bibnamefont{Jona}},
  \bibinfo{journal}{Europ. Phys. J. B} \textbf{\bibinfo{volume}{26}},
  \bibinfo{pages}{81} (\bibinfo{year}{2002}).

\bibitem[{\citenamefont{Reichert and Stark}(2005)}]{reichert}
\bibinfo{author}{\bibfnamefont{M.}~\bibnamefont{Reichert}} \bibnamefont{and}
  \bibinfo{author}{\bibfnamefont{H.}~\bibnamefont{Stark}},
  \bibinfo{journal}{Eur.\ Phys.\ J.\ E} \textbf{\bibinfo{volume}{17}},
  \bibinfo{pages}{493} (\bibinfo{year}{2005}).

\bibitem[{\citenamefont{Lenz and Ryskin}(2006)}]{lenz06}
\bibinfo{author}{\bibfnamefont{P.}~\bibnamefont{Lenz}} \bibnamefont{and}
  \bibinfo{author}{\bibfnamefont{A.}~\bibnamefont{Ryskin}},
  \bibinfo{journal}{Phys. Biol.} \textbf{\bibinfo{volume}{3}},
  \bibinfo{pages}{285} (\bibinfo{year}{2006}).

\bibitem[{\citenamefont{Vilfan and Julicher}(2006)}]{vilfan}
\bibinfo{author}{\bibfnamefont{A.}~\bibnamefont{Vilfan}} \bibnamefont{and}
  \bibinfo{author}{\bibfnamefont{F.}~\bibnamefont{Julicher}},
  \bibinfo{journal}{Phys. Rev. Lett.} \textbf{\bibinfo{volume}{96}},
  \bibinfo{pages}{058102} (\bibinfo{year}{2006}).

\bibitem[{\citenamefont{Guirao and Joanny}(2007)}]{guirao07}
\bibinfo{author}{\bibfnamefont{B.}~\bibnamefont{Guirao}} \bibnamefont{and}
  \bibinfo{author}{\bibfnamefont{J.~F.} \bibnamefont{Joanny}},
  \bibinfo{journal}{Biophys. J.} \textbf{\bibinfo{volume}{92}},
  \bibinfo{pages}{1900} (\bibinfo{year}{2007}).

\bibitem[{\citenamefont{Niedermayer et~al.}(2008)\citenamefont{Niedermayer,
  Eckhardt, and Lenz}}]{niedermayer08}
\bibinfo{author}{\bibfnamefont{T.}~\bibnamefont{Niedermayer}},
  \bibinfo{author}{\bibfnamefont{B.}~\bibnamefont{Eckhardt}}, \bibnamefont{and}
  \bibinfo{author}{\bibfnamefont{P.}~\bibnamefont{Lenz}},
  \bibinfo{journal}{Chaos} \textbf{\bibinfo{volume}{18}},
  \bibinfo{pages}{037128} (\bibinfo{year}{2008}).

\bibitem[{\citenamefont{Elfring and Lauga}(2009)}]{ElfringLauga2009}
\bibinfo{author}{\bibfnamefont{G.~J.} \bibnamefont{Elfring}} \bibnamefont{and}
  \bibinfo{author}{\bibfnamefont{E.}~\bibnamefont{Lauga}},
  \bibinfo{journal}{Phys. Rev. Lett.} \textbf{\bibinfo{volume}{103}},
  \bibinfo{pages}{088101} (\bibinfo{year}{2009}).

\bibitem[{\citenamefont{Kim and Powers}(2004)}]{kim}
\bibinfo{author}{\bibfnamefont{M.~J.} \bibnamefont{Kim}} \bibnamefont{and}
  \bibinfo{author}{\bibfnamefont{T.~R.} \bibnamefont{Powers}},
  \bibinfo{journal}{Phys.\ Rev.\ E} \textbf{\bibinfo{volume}{69}},
  \bibinfo{pages}{061910} (\bibinfo{year}{2004}).

\bibitem[{\citenamefont{Alexander and Yeomans}(2008)}]{alexander08}
\bibinfo{author}{\bibfnamefont{G.~P.} \bibnamefont{Alexander}}
  \bibnamefont{and} \bibinfo{author}{\bibfnamefont{J.~M.}
  \bibnamefont{Yeomans}}, \bibinfo{journal}{Euro. Phys. Lett.}
  \textbf{\bibinfo{volume}{83}}, \bibinfo{pages}{34006} (\bibinfo{year}{2008}).

\bibitem[{\citenamefont{Lauga and Bartolo}(2008)}]{laugabartolo08}
\bibinfo{author}{\bibfnamefont{E.}~\bibnamefont{Lauga}} \bibnamefont{and}
  \bibinfo{author}{\bibfnamefont{D.}~\bibnamefont{Bartolo}},
  \bibinfo{journal}{Phys. Rev. E} \textbf{\bibinfo{volume}{78}},
  \bibinfo{pages}{030901} (\bibinfo{year}{2008}).

\bibitem[{\citenamefont{Bennett et~al.}(2002)\citenamefont{Bennett, Schatz,
  Rockwood, and Wiesenfeld}}]{Huygenss-Clocks}
\bibinfo{author}{\bibfnamefont{M.}~\bibnamefont{Bennett}},
  \bibinfo{author}{\bibfnamefont{M.~F.} \bibnamefont{Schatz}},
  \bibinfo{author}{\bibfnamefont{H.}~\bibnamefont{Rockwood}}, \bibnamefont{and}
  \bibinfo{author}{\bibfnamefont{K.}~\bibnamefont{Wiesenfeld}},
  \bibinfo{journal}{Proc. R. Soc. Lond. A} \textbf{\bibinfo{volume}{458}},
  \bibinfo{pages}{563} (\bibinfo{year}{2002}).

\bibitem[{\citenamefont{Cortez}(2001)}]{Cortez2001}
\bibinfo{author}{\bibfnamefont{S.}~\bibnamefont{Cortez}},
  \bibinfo{journal}{{SIAM} J. Sci. Comput.} \textbf{\bibinfo{volume}{{\bf
  23}}}, \bibinfo{pages}{1204} (\bibinfo{year}{2001}).

\bibitem[{\citenamefont{Doi and Edwards}(1986)}]{doi_edwards1986}
\bibinfo{author}{\bibfnamefont{M.}~\bibnamefont{Doi}} \bibnamefont{and}
  \bibinfo{author}{\bibfnamefont{S.}~\bibnamefont{Edwards}},
  \emph{\bibinfo{title}{The theory of polymer dynamics}}
  (\bibinfo{publisher}{Oxford University Press}, \bibinfo{address}{Oxford},
  \bibinfo{year}{1986}).

\bibitem[{\citenamefont{Strogatz}(1994)}]{Strogatz1994}
\bibinfo{author}{\bibfnamefont{S.~H.} \bibnamefont{Strogatz}},
  \emph{\bibinfo{title}{Nonlinear dynamics and chaos}}
  (\bibinfo{publisher}{Perseus Books}, \bibinfo{address}{Reading, MA},
  \bibinfo{year}{1994}).

\bibitem[{\citenamefont{Russel et~al.}(1989)\citenamefont{Russel, Saville, and
  Schowalter}}]{russel_etal1989}
\bibinfo{author}{\bibfnamefont{W.}~\bibnamefont{Russel}},
  \bibinfo{author}{\bibfnamefont{D.}~\bibnamefont{Saville}}, \bibnamefont{and}
  \bibinfo{author}{\bibfnamefont{W.}~\bibnamefont{Schowalter}},
  \emph{\bibinfo{title}{Colloidal dispersions}} (\bibinfo{publisher}{Cambridge
  University Press}, \bibinfo{address}{Cambridge}, \bibinfo{year}{1989}).

\end{thebibliography}

\end{document}